\documentclass[aip,amsmath,amssymb,jcp,nofootinbib,12pt
]{revtex4-1}
\usepackage[utf8]{inputenc}
\usepackage{amsfonts}
\usepackage{amsmath}
\usepackage{amssymb}
\usepackage{amsthm}
\usepackage{appendix}
\usepackage{bbm}
\usepackage{chemformula}
\usepackage{enumerate}
\usepackage[T1]{fontenc}
\usepackage{datetime}
\usepackage{dcolumn} 
\usepackage{epigraph}
\usepackage[T1]{fontenc}
\usepackage{graphicx} 
\usepackage{hyperref}
\hypersetup{
    colorlinks=true,
    linkcolor=blue,
    filecolor=magenta,
    urlcolor=cyan,
}
\usepackage{cleveref}
\usepackage{soul} 
\usepackage{tabularx}
\usepackage{xcolor}
\usepackage{yfonts}

\newcommand{\op}[1]{\mathsf #1}

\renewcommand*{\bar}{\thickbar}
\newcommand{\erf}{\mathrm{erf}}
\newcommand{\erfc}{\mathrm{erfc}}
\newcommand{\bra}[1]{\ensuremath{\langle #1 \vert}}
\newcommand{\ket}[1]{\ensuremath{\vert #1  \rangle}}

\DeclareRobustCommand{\varlambda}{\text{\usefont{OML}{txmi}{m}{it}\symbol{"15}}}

\newcommand*\dd{\mathop{}\!\mathrm{d}}

\newcommand{\bfr}{{\bf r}}

\makeatletter
\newcommand{\thickbar}{\mathpalette\@thickbar}
\newcommand{\@thickbar}[2]{{#1\mkern1.5mu\vbox{
  \sbox\z@{$#1\mkern-1.5mu#2\mkern-1.5mu$}%
  \sbox\tw@{$#1\overline{#2}$}%
  \dimen@=\dimexpr\ht\tw@-\ht\z@-.8\p@\relax
  \hrule\@height.8\p@ 
  \vskip\dimen@
  \box\z@}\mkern1.5mu}
}
\makeatother

\begin{document}
\title{Second-order adiabatic connection.\\ 
The theory and application to two electrons in a parabolic confinement.}
\author{Andreas Savin$^\ast$}
\affiliation{Laboratoire de Chimie Théorique, CNRS and Sorbonne University,
4 place Jussieu, 75252 Paris, France}
\email{andreas.savin@lct.jussieu.fr}
\author{Jacek Karwowski}
\affiliation{Institute of Physics, Faculty of Physics, Astronomy and 
Informatics, Nicolaus Copernicus University, Grudziadzka 5, 87-100 Toru\'n, 
Poland}
\email{jka@umk.pl}

\ddmmyyyydate
\today \; \currenttime

\begin{quote}
 This article has been published in \\ The Journal of Chemical Physics {\bf 159}, 134107 (2023) \\
\url{https://doi.org/10.1063/5.0167851}
\end{quote}

\begin{abstract}
The adiabatic connection formalism, usually based on the first-order
perturbation theory, has been generalized to an arbitrary order. The
generalization stems from the observation that the formalism can be
derived from a properly arranged Taylor expansion. The second-order theory
is developed in detail and applied to the description of two electrons in a
parabolic confinement (harmonium). A significant improvement relative to the 
first-order theory has been obtained.  
\end{abstract}

\maketitle

\tableofcontents

\section{Introduction}
The adiabatic connection is used in quantum mechanics to express corrections
to models by progressively approaching the system of interest.~\cite{Gut-31}
Usually, this is formally obtained by using for each infinitesimal step
the first-order perturbation theory.  This paper generalizes the idea of
adiabatic connection (as used in quantum mechanics) by applying it at higher
orders of perturbation theory [see eq. \eqref{eq:working-2}]. 
Mathematically, it corresponds to using the integral remainder in Taylor's
formula.  We thus expect improvement by going to higher orders.

The advantage of such an approach originates from using operators that require
reduced information about the wave function.  In our case, we exploit the
short-range behavior of the wave function.~\cite{Sav-JCP-20} As the short-range
behavior has features independent of a specific (electronic) systems, it can
be applied "universally", that is, in a system-independent way, in the
spirit of density functional theory (DFT).  The approach retains the
fundamental role of the adiabatic connection in DFT where it was used not
only for explaining what the exchange-correlation density functional should
do~\cite{HarJon-JPF-74, LanPer-SSC-75, GunLun-PRB-76, ZieRauBae-TCA-77}, but
also as a guide in constructing density functional approximations (see, {\em
e.g.}, refs.~\onlinecite{Bec-IJQC-83, ErnPer-JCP-98}).  As in DFT, we need
to complement by information provided by the model system.  Our approach
avoids certain of the limitations present in density functional theory: it
is valid for any state, and it needs no fitting to systems such as the
uniform electron gas.  No use of the Hohenberg-Kohn
theorem~\cite{HohKoh-PR-64}, is made.  Thus, the method presented here is
not restricted to ground states.

Although generally applicable, we illustrate our method by applying it to a
system of two electrons in a parabolic confinement (harmonium), as it is
sufficient to illustrate the aspects mentioned above.

\section{Theory}
\subsection{Adiabatic connection}
Mathematically, the idea of adiabatic connection relies on the equation
\begin{equation}
    \label{eq:ac1-math}
    f(b) = f(a) + \int_a^b  \dd \lambda \: f'(\lambda)
\end{equation}
where $f'(\lambda)$ is the first derivative of the function $f(\lambda)$. 
In quantum mechanics, the function $f(\lambda)$ is often associated to an
energy.  For this, let us consider some model Hamiltonian, $H(\mu)$, and the
corresponding Schr\"odinger equation,
\begin{equation}
    \label{eq:se-model}
    H(\mu) \Psi(\mu) = E(\mu) \Psi(\mu).
\end{equation}
Note that to simplify the notation, we drop the designations of
coordinates and quantum numbers whenever this does not lead to
misunderstandings.
The quantity of interest is not the energy of the model system, $E(\mu)$,
but $E$, the energy produced by the Schr\"odinger equation of the physical
Hamiltonian, $H$.  We write
\begin{equation}
    \label{eq:h-lambda}
    H(\lambda,\mu) = H(\mu) + \lambda \left( H-H(\mu)\right)
\end{equation}
and
\begin{equation}
    \label{eq:se-lambda}
    H(\lambda,\mu) \Psi(\lambda,\mu) = E(\lambda,\mu) \Psi(\lambda,\mu)
\end{equation}
Note that for $\lambda=0$ we have the model system, and for $\lambda=1$ the
physical system.  Furthermore (by first-order perturbation theory, the
Hellmann-Feynman theorem),
\begin{equation}
    \label{eq:hel-fey}
    \partial_\lambda E(\lambda,\mu) = \bra{\Psi(\lambda, \mu)} H-H(\mu) 
    \ket{{\Psi(\lambda, \mu)}}
\end{equation}
Applying eq.~\eqref{eq:ac1-math}, we have an expression for $E(\lambda,\mu)$, 
\begin{equation}
    \label{eq:e-lambda}
    E(\lambda,\mu) = E(\mu) + \int_0^\lambda \dd \varlambda \, 
    \bra{\Psi(\varlambda, \mu)} H-H(\mu) \ket{{\Psi(\varlambda, \mu)}}.
\end{equation}
In particular, for $\lambda=1$, we have the correction that added to
$E(\mu)$ produces the physical energy, $E$
\begin{align}
    \label{eq:e-bar}
    \bar{E}(\mu) & = E - E(\mu) \\
    \label{eq:ac-1}
            & = \int_0^1 \dd \lambda \, \bra{\Psi(\lambda, \mu)} H-H(\mu) 
                \ket{{\Psi(\lambda, \mu)}}
\end{align}
Eq.~\eqref{eq:ac-1} seems useless, as it requires the knowledge of the wave
function for $\lambda \in [0,1]$.  However, one can exploit
eq.~\eqref{eq:ac-1}, if the behavior of these wave functions is known for
the specific domain probed by the operator $H-H(\mu)$.  In this paper we
consider
\begin{equation}
\label{eq:this-H}
H(\mu) = T + V + W(\mu)
\end{equation}
where $T$ is the kinetic energy operator, $V=\sum_{i=1}^N v(\bfr_i)$, and
$W(\mu)=\sum_{i<j} w(|\bfr_i - \bfr_j|,\mu)$ are one- and two-particle
potential energy operators.  Note that only  the two-particle operator is
model-dependent.  We choose $w(r;\mu)$ as a long-range operator, in order to
have
\begin{equation}
    \label{eq:W-bar}
    \bar{W}(\mu) = H - H(\mu),
\end{equation}
a short-range operator to show up in eq.~\eqref{eq:ac-1}, and exploit the
short-range properties of the wave function.  All the numerical examples
below are produced with
\begin{align}
    \label{eq:w}
    w(r;\mu) & = \frac{\erf(\mu r)}{r} \\
    \label{eq:w-bar}
    \bar{w}(r;\mu) & = \frac{\erfc(\mu r)}{r}
\end{align}

\subsection{Second-order adiabatic connection}
Eq.~\eqref{eq:ac1-math} is only a particular case of the Taylor's expansion
with the integral form of the remainder,

\begin{equation}
    \label{eq:ac-k-math}
    f(b) = \left( \sum_{k=0}^{K-1} \frac{1}{k!}(b-a)^k f^{(k)}(a) \right) 
    + \int_a^b \dd\lambda \, \frac{1}{(K-1)!} (b-\lambda)^{K-1} f^{(K)}(\lambda)
\end{equation}
where $f^{(k)}$ is the $k^\text{th}$ derivative of $f$; 
eq.~\eqref{eq:ac1-math} is obtained for $K=1$.
We consider below the case $K=2$,
\begin{equation}
    \label{eq:ac-2}
    E = E(\mu) + \left. \partial_\lambda E(\lambda,\mu) \right|_{\lambda=0}  
    + \int_0^1 \dd \lambda \, (1-\lambda) \, \partial_\lambda^2 E(\lambda,\mu) 
\end{equation}
In order to make the distinction between the variants of adiabatic
connection, we call the usual one, eq.~\eqref{eq:ac-1} {\em first-order}
adiabatic connection, AC1, and that given by eq.~\eqref{eq:ac-2} {\em
second-order} adiabatic connection, AC2.  Using higher-order adiabatic
connections is possible, but they are not explored in this paper.  We would
like to stress that eqs.~\eqref{eq:ac-1} and \eqref{eq:ac-2} are both
rigorous, as long as the derivatives exist.

Note that neglecting the integral on the right-hand side of
eq.~\eqref{eq:ac-2} gives the first-order perturbation theory expression,
and making the approximation $\partial_\lambda^2 E(\lambda,\mu) \approx
\partial_\lambda^2 E(\lambda,\mu)|_{\lambda=0}$ gives the second-order
perturbation theory expression,
\begin{equation}
    \label{eq:pt-2}
    E \approx E(\mu) + \left. \partial_\lambda E(\lambda,\mu) \right|_{\lambda=0}  
    + \left. \frac{1}{2} \partial_\lambda^2 E(\lambda,\mu)\right|_{\lambda=0}
\end{equation}
Using $\left.  \partial_\lambda E(\lambda,\mu) \right|_{\lambda=0}$ in
eq.~\eqref{eq:ac-2} requires only using eq.~\eqref{eq:hel-fey} at
$\lambda=0$, that is, the model wave function, $\Psi(\mu)$,
\begin{equation}
    \label{eq:dedlambda}
    \partial_\lambda E(\lambda,\mu) =  \bra{\Psi(\mu)} \bar{W}(\mu) \ket{\Psi(\mu)} 
\end{equation}
For obtaining the second-derivative in eq.~\eqref{eq:ac-2}, we use eq.~\eqref{eq:e-lambda},
\begin{equation}
    \label{eq:d2edlambda2}
    \partial_\lambda^2 E(\lambda,\mu) = \partial_\lambda \bra{\Psi(\lambda,\mu)} \bar{W}(\mu) 
    \ket{\Psi(\lambda,\mu)}
\end{equation}

\subsection{Exploiting the short-range behavior of the wave function}

In order to avoid the need of knowing $\Psi$ for all $\lambda$ in
eqs.~\eqref{eq:dedlambda} and \eqref{eq:d2edlambda2}, we use a short-range
operator $\bar{W}$, and exploit the "universal" features of the wave
function.
The behavior of the model wave function approaching the physical system can
be analyzed by considering the limit of large $\mu$.~\cite{GorSav-PRA-06} In
this limit, the difference between the model and physical wave function is
dominated by the behavior at short electron-electron distances (as
$\bar{w}(r;\mu)$ differs from the Coulomb potential only in this domain). 
For $r=|\bfr_1 - \bfr_2| \rightarrow 0$,
\begin{equation}
\label{eq:psi-sr}
\Psi(\lambda, \mu) = \sum_\ell \mathcal{N}_\ell \, \varphi_\ell(r;\lambda,\mu),
\end{equation}
where $\ell$ is an angular quantum number related to
$\bfr_1-\bfr_2$, and 
$\mathcal{N}_\ell$ depends on all variables except  $r$.

Below, we use $\varphi_\ell(r;\lambda,\mu)$ that is correct to order
$\mu^{-1}$.  It satisfies the generalization of Kato's cusp condition to
$w(r;\mu) + \lambda \bar{w}(r;\mu)$.~\cite{Kat-CPAM-57, PacBro-JCP-66,SilUgaBoy-00,
KurNakNak-AQC-16}.  Its explicit form is derived in Appendix:
\begin{align}
\label{eq:wf-asy}
\varphi_\ell(r;\lambda,\mu)\,\propto\,&r^\ell\left[1+\frac{\lambda\,r}{2\ell+2}
+\frac{1-\lambda}{2\ell+2}\left(r\,\erf(\mu r)+\frac{2\ell+2}{2\ell+1} 
\frac{e^{-\mu^2 r^2}}{\mu\sqrt{\pi}}\right.\right.\\
& \nonumber 
+\left.\left.\frac{\Gamma(\ell+3/2)-\Gamma(\ell+3/2,\mu^2r^2)}
{\sqrt{\pi}(2\ell+1)\mu^{2\ell+2}r^{2\ell+1}}\right)\right]
\end{align}
where the incomplete $\Gamma$ function is:
\[ \Gamma(a,z) = \int_z^\infty \dd t \, t^{a-1} e^{-t} \]
and $\Gamma(a)=\Gamma(a,0)$.

In most quantum-chemical methods one considers the natural singlet and triplet
pairing, corresponding to $\ell=0$, and $\ell=1$, respectively.  The
{\em non-natural} singlets and triplets, as they are called by Kutzelnigg and Morgan~
\cite{KutMor-JCP-92}, correspond to $\ell\,>\,1$.
If $\ell\,>\,1$ then the centrifugal force keeps
electrons apart. As a consequence, for $r\,\rightarrow\,0$ the 
prefactor $r^\ell$ in eq. \eqref{eq:wf-asy} decreases with increasing $\ell$.
Consequently, the terms with small $\ell$ dominate in 
expansion~\eqref{eq:psi-sr}. 
For the system treated in this paper (harmonium), due to separability, the sum is exactly
reduced to one term and, in principle, one can study the behavior of the
wave function corresponding to an arbitrary $\ell$.
In the remaining part of this Section the
$\ell$-dependence is marked explicitly: $\Psi(\lambda,\mu)$ is denoted as
$\Psi_\ell(\lambda,\mu)$, and $E(\lambda,\mu)$ - as $E_\ell(\lambda,\mu)$.

\subsection{The model system}
All numerical results presented hereafter are obtained for a system of two
electrons in a parabolic confinement (harmonium) with $v=\frac{1}{2}
\omega^2 r^2$.  If not specified otherwise, $\omega=1/2$~a.u.  The
interaction between electrons is generalized to $w$, eq.~\eqref{eq:w}.  The
Schr\"odinger equation is separable, and we have to consider only the
one-dimensional radial equation
\begin{equation}
    \label{eq:harmonium}
    \left( -\partial_r^2 -\frac{2}{r} \partial_r + 
    \frac{\ell (\ell+1)}{r^2} + \frac{1}{4} \omega^2 r^2 + w(r;\mu) + 
    \lambda \bar{w}(r;\mu)-E_\ell(\lambda,\mu) \right) 
    \varphi_\ell(r;\lambda,\mu)=0
\end{equation}
It can be solved to arbitrary accuracy, and this allows us to judge the
errors made to approximations.  Furthermore, there is no need to take into
account the other coordinates in the prefactor $\mathcal{N}_\ell$ 
showing up in eq.~\eqref{eq:psi-sr}.

\subsection{Working equations}
In first-order adiabatic connection we approximate
\begin{equation}
  \label{eq:approx-pt1}
   \partial_\lambda E_\ell(\lambda,\mu) = \bra{\Psi(\lambda,\mu)} 
   \bar{W}(\mu) \ket{\Psi(\lambda,\mu)} \approx c_\ell \,\mathcal{I}_\ell^{(1)}(\lambda,\mu),
\end{equation}
where $c_\ell=\mathcal{N}_\ell^2$ and
\begin{equation}
    \label{eq:i1}
    \mathcal{I}^{(1)}_\ell(\lambda,\mu) = \int_0^\infty \dd r \, r^2 \, 
    |\varphi_\ell(r; \lambda, \mu)|^2 \bar{w}(r;\mu).
\end{equation}
In second-order adiabatic connection we approximate
\begin{equation}
  \label{eq:approx-pt2}
   \partial_\lambda^2 E_\ell(\lambda,\mu) \approx c_\ell \, 
  \mathcal{I}^{(2)}_\ell(\lambda,\mu)
\end{equation}
where 
\begin{equation}
    \label{eq:i2}
    \mathcal{I}^{(2)}_\ell(\lambda,\mu) = 
    \partial_\lambda \mathcal{I}^{(1)}_\ell(\lambda,\mu).
\end{equation}
As $\varphi_\ell(r;\lambda,\mu)$ is explicitly known (in the given limit: $r
\rightarrow 0, \mu \rightarrow \infty$), the integrals
$\mathcal{I}_\ell^{K}$ are computed to be (for $\ell=0$ and $\ell=1$),
\begin{align}
    \label{eq:i1-l0}
    \mathcal{I}_{\ell=0}^{(1)} & \propto \frac{1}{\mu^2}+\frac{a_1 + a_2
    (1-\lambda)}{\mu^3} +\frac{a_3 + a_4 (1 - \lambda) + a_5 (1 -
    \lambda)^2}{\mu^4} \\
    \label{eq:i1-l1}
     \mathcal{I}_{\ell=1}^{(1)} & \propto \frac{1}{\mu^4} + \frac{b_1 + b_2
     (1 - \lambda)}{\mu^5} + \frac{b_3 + b_4 (1 - \lambda) + b_5 (1 -
     \lambda)^2}{\mu^6}
\end{align}
where the constants $a_k$ and $b_k$ are collected in table~\ref{tab:ak}.

\begin{table}[h]
\begin{tabular}{c|ccccc}
\hline
$k$    & 1          &2           &3           &4           &5\\  \hline
$a_k$\;&\; 0.75225\;&\; 0.62319\;&\; 0.18750\;&\; 0.19331 \;&\; 0.10700\\
$b_k$\;&\; 0.60180\;&\; 0.07301\;&\; 0.10417\;&\; 0.01647 \;&\; 0.00198\\
\hline
\end{tabular}
\caption{Constants in eqs.~\eqref{eq:i1-l0} and \eqref{eq:i1-l1} 
rounded up to five decimals.\label{tab:ak}} 
\end{table}

The proportionality constant, $c$, is related to the physical system and can
be determined by using eqs.~\eqref{eq:approx-pt1} and \eqref{eq:approx-pt2}
for the model system (at $\lambda=0$).  We thus get the following working
equations:
\begin{equation}
    \label{eq:working-1}
    E_\ell \approx E_\ell(\mu) + \alpha_\ell^{(1)}(\mu) \left. 
    \partial_\lambda E_\ell(\lambda,\mu) \right|_{\lambda=0}
\end{equation}
and 
\begin{equation}
    \label{eq:working-2}
        E_\ell \approx E_\ell(\mu) + 
    \left. \partial_\lambda E_\ell(\lambda,\mu) 
        \right|_{\lambda=0} + \alpha_\ell^{(2)}(\mu) \left. 
        \partial_\lambda^2 E_\ell(\lambda,\mu) \right|_{\lambda=0}
\end{equation}
with
\begin{equation}
    \label{eq:alpha-K}
    \alpha_\ell^{(K)}(\mu) = \frac{\int_0^1 \dd \lambda \, 
    (1-\lambda)^{(K-1)} \, \mathcal{I}^{(K)}_\ell(\lambda,\mu)}
    {\mathcal{I}^{(K)}_\ell(\lambda=0,\mu)} \\
\end{equation}
The integrals over $\lambda$ are trivial, and not shown. 
Eq.~\eqref{eq:working-1} corresponds to the first-order adiabatic
connection, while eq.~\eqref{eq:working-2} corresponds to second-order
adiabatic connection.  Note that expressions~\eqref{eq:working-1} and
\eqref{eq:working-2} require the same effort as the first- and second-order
perturbation theory, respectively, namely computing $\partial_\lambda
E_\ell(\lambda,\mu)_{\lambda=0}$ and $\partial_\lambda^2 E_\ell(\lambda,\mu)
|_{\lambda=0}$.  Only the weight of the last term is changed by
$\alpha_\ell^{(K)}$, $K=1\,\text{or}\,2$.

Two more remarks on these equations.  First, note that by squaring
$\varphi_\ell$ in eq.~\eqref{eq:i1} we introduce terms in $\mu^{-2}$,
although a further term to this order may be present in an exact theory,
because $\varphi_\ell$ is constructed to order $\mu^{-1}$ only.  Second,
higher orders in the adiabatic connection do not improve over perturbation
theory with the present form of $\varphi_\ell$: in our first-order
expression in $1/\mu$, only terms linear in $\lambda$ show up.  The second
derivative is just a constant, and $\alpha_\ell^{(3)}=1$.

Eq.~\eqref{eq:working-1}  can be rewritten as 
\begin{equation}
 \label{eq:working-1-corr}
 E_\ell \approx \bra{\Psi(\mu)} H \ket{\Psi(\mu)} + \left(\alpha_\ell^{(1)}(\mu) -1 \right) \left. 
    \partial_\lambda E_\ell(\lambda,\mu) \right|_{\lambda=0}
\end{equation}
The last term on the r.h.s. appears as approximation for the correlation energy.
One can also write eq.~~\eqref{eq:working-2}, as
\begin{equation}
 \label{eq:working-2-corr}
 E_\ell \approx \bra{\Psi(\mu)} H \ket{\Psi(\mu)} +
 E_\ell^{(2)}(\mu) + \left(\alpha_\ell^{(2)}(\mu) -\frac{1}{2} \right) \left. 
    \partial_\lambda^2 E_\ell(\lambda,\mu) \right|_{\lambda=0}
\end{equation}
where $E_\ell^{(2)}(\mu)=1/2 \, \partial_\lambda^2
E_\ell(\lambda,\mu)|_{\lambda=0}$ is the second-order energy correction to
$E_\ell(\mu)$.  The last term on the r.h.s.  appears as approximation for
the remaining correlation energy.

\section{Numerical results}
\subsection{General considerations}
By construction, the approximations leading to the working
equations~\eqref{eq:working-1} and \eqref{eq:working-2} become exact in the
limit of the model system approaching the physical system ($\mu \rightarrow
\infty$).  However, the cost of performing the calculation of the model
system also increases with $\mu$.  We are thus interested to use models with
small $\mu$, ideally even totally turn off the interaction ($\mu = 0$).  The
derivation does not tell us how well the approximations used work for small
$\mu$.  The figures shown in this paper present the errors of the
approximation (the difference between the energy obtained using it and that
of the physical system) for different models $\mu$.  As $\mu$ can vary
continuously, the plots of the errors show up as curves.  We consider $\mu
\in [0,3]$, as $E(\mu)$ approaches anyhow the physical energy $E$ for large
$\mu$.

With these approximations, we aim to reach the so-called {\em chemical
accuracy}, i.e., 1~kcal/mol.~\cite{Pop-RMP-99} The region of chemical
accuracy is marked in the plots by horizontal dashed lines.

\subsection{Discussion of the figures}
\begin{figure}
    \centering
    \includegraphics{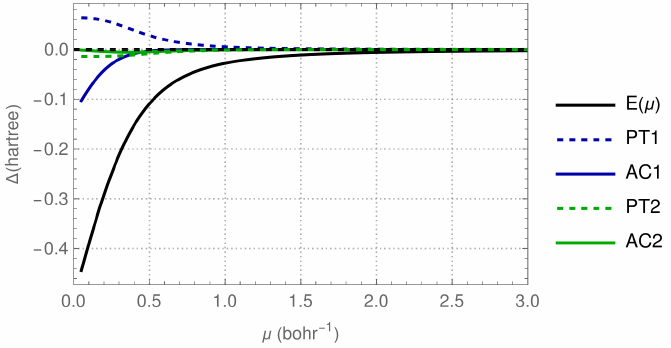}
    \caption{Errors of different approximations for the two-electron
harmonium ($\omega=1/2$).  The full black curve indicates the error of the
model, $E(\mu)$.  Blue curves correspond to first-order: dashed curve for
perturbation theory, $\bra{\Psi(\mu)} H \ket{\Psi(\mu)}$, full curve for
first-order adiabatic connection, eq.~\eqref{eq:working-1}.  Green curves
correspond to second-order: dashed curve for perturbation theory,
eq.~\eqref{eq:pt-2}, full curve for second-order adiabatic connection,
eq.~\eqref{eq:working-2}.}
\label{fig:paolasall}
\end{figure}
Fig.~\ref{fig:paolasall} shows the general trends of the approximations.  One
can see, that by the choice of using the bare field, $v$, for all values of
$\mu$, cf.  eq.~\eqref{eq:this-H}, the error of the model in the limit of
small $\mu$ is catastrophic (0.5~hartree).  First-order perturbation theory
leads to the expectation value of the physical Hamiltonian, and thus gives an
upper bound to the exact energy.  In spite of using a bad wave function, the
improvement is important: the error is decreased by an order of magnitude to
$\approx 0.06$~hartree.  Hartree-Fock, the best value that can be obtained
for the non-interacting wave function, is in error by $\approx 0.04$~hartree. 
Second-order perturbation theory further improves the result.

\begin{figure}
    \centering
    \includegraphics{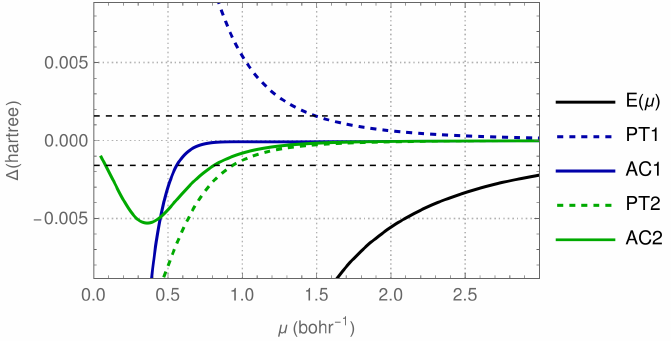} \\
    \includegraphics{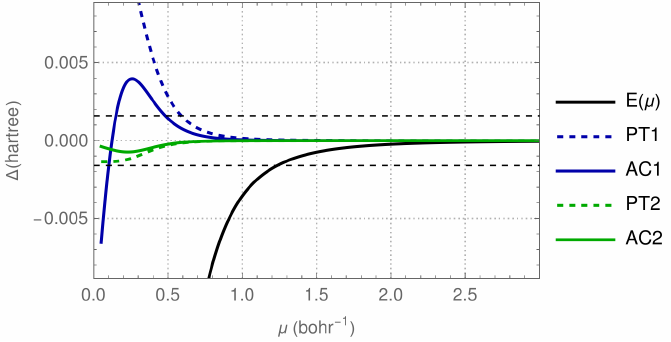} \\
    \includegraphics{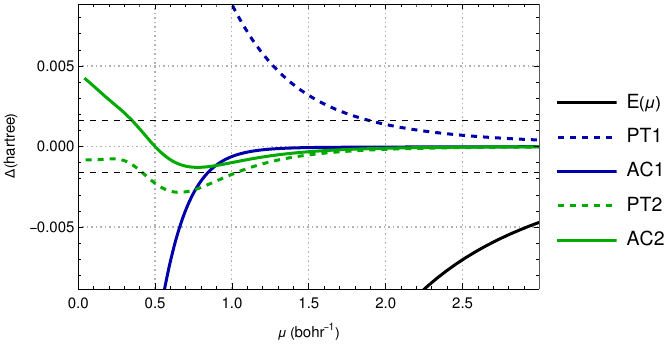}
    \caption{Errors of different approximations for the two-electron
harmonium ($\omega=1/2$).  The full black curve indicates the error of the
model, $E(\mu)$.  Blue curves correspond to first-order: dashed curve for
perturbation theory, $\bra{\Psi(\mu)} H \ket{\Psi(\mu)}$, full curve for
first-order adiabatic connection, eq.~\eqref{eq:working-1}.  Green curves
correspond to second-order: dashed curve for perturbation theory,
eq.~\eqref{eq:pt-2}, full curve for second-order adiabatic connection,
eq.~\eqref{eq:working-2}.  Top panel: ground state; middle panel: first
excited state with $\ell=1$; bottom panel: first excited state with
$\ell=0$.}
\label{fig:paolas5}
\end{figure}
The range of chemical accuracy cannot be distinguished on the scale of the
plot shown in fig.~\ref{fig:paolasall}.  In order to discuss the suitability
of the results, we have to zoom in (see fig.~\ref{fig:paolas5}, top).  We
notice that for the ground state (a singlet state, $\ell=0$), the model does
not reach chemical accuracy for the whole range of $\mu$ shown in the plot. 
The result of the first-order perturbation theory is improved by using the
correction established for $\mu \rightarrow \infty$,
eq.~\eqref{eq:working-1}.  Surprisingly, in spite of using an expansion in
$1/\mu$, the error remains negligible even for some range of $\mu <1
$~bohr$^{-1}$ (to $\approx 0.5$~bohr$^{-1}$).  However, at smaller $\mu$ it
worsens dramatically.  Going over to the second-order perturbation theory
improves over the first-order perturbation theory.  Correcting asymptotically,
eq.~\eqref{eq:working-2} improves the result over the whole range of values. 
However, chemical accuracy is not yet reached for all models.

Let us now apply the method to excited states.  The model works much better
for first triplet excited state ($\ell=1$), see fig.~\ref{fig:paolas5}
(middle panel).  This improves the results for all approximations.

Considering the first singlet excited state with $\ell=0$,
fig.~\ref{fig:paolas5} (bottom), we notice that the quality of the model is
worse than for the ground state.  The corresponding curve does not even show
up in the plot.  However, the approximations show a behavior that roughly
follows that of the ground state.  We would like to stress that the
corrections for the ground state and for the first excited state with
$\ell=0$ use the same factors $\alpha_\ell^{(K)}$, eq.  (29).  The
corrections are, of course, different, as the system-specific information
enters through the model-specific quantities, 
$\partial^K E_\ell(\lambda,\mu)|_{\lambda=0}$.

\section{Conclusion}

The integral form of the remainder in Taylor's expansion,
eq.~\eqref{eq:ac-k-math} provides a formula that generalizes the adiabatic
connection.  We use it to construct approximations to correct energies
produced by model Hamiltonians with long range interaction,
eqs.~\eqref{eq:working-1} and \eqref{eq:working-2}.  They are inspired by
approximations used in density functional theory.  However, they do not use
the Hohenberg-Kohn theorem, and are valid for the ground and excited states. 
Instead, they are constructed to be valid for the short-range interaction,
as one approaches the physical system.
 
Results are shown only for harmonium.  One can notice an improvement with
increasing order of the adiabatic connection.  No comparison is made with
analogous density functional approximations.  These can be found in
ref.~\onlinecite{Sav-JCP-20}. 

Of course, one would like to apply the method not only to harmonium.
In other systems the sum in eq.~\eqref{eq:psi-sr} is extended over
all values of $\ell$. An explicit treatment of the higher terms 
is difficult, may be too expensive computationally, and is maybe not needed -- 
as it was alredy discussed, these terms are usually much smaller than the
leading one.
Thus, techniques such as described in refs.~\onlinecite{McWKut-IJQC-68}, and
\onlinecite{Ten-JCP-04} could be applied.
 
A strength of the method presented here is the stability of
the result for large $\mu$.  Once the stability is lost, it may
indicate a worsening of the approximation.

A weakness of the method presented here is that it does not work
sufficiently well (within chemical accuracy) for the non-interacting system
($\mu=0$).  This makes the method more expensive.  This feature is also
present in range-separated density functional methods which possess a
striking similarity with the one presented here.  One may wonder if the
experience gained by constructing density functional approximations cannot
be exploited here, too.  In particular, a properly constructed effective
one-body potential in the model Hamiltonian can reduce the energy error of
the physical (interacting) system (see, e.g., figs.~11 and 12 in
ref.~\onlinecite{Sav-JCP-20}).  Alternatively, the method presented here
could be used to improve density functional approximations.

Finally, we would like to point out that eq.~\eqref{eq:working-2} can be
seen as a theoretical justification for spin-component-scaled
methods~\cite{GriGoeFin-12}.

\section{Acknowledgement}
We dedicate this paper to John P.  Perdew who transformed the adiabatic
connection to a useful tool not only in understanding DFT, but also for
constructing approximations that changed the impact of computational
chemistry and physics.

\appendix*
\section{A derivation of equation (19)}

We generalize for an arbitrary $\ell$ the results obtained for $\ell=0$ in
Section III of ref.\citenum{GorSav-PRA-06}, and in Appendix of
ref.\citenum{SavKar-23}.  We consider the behavior at the limit of large
$\mu$ of the radial wave function $\varphi(r;\lambda,\mu)$ which is a
solution of the radial Schr\"odinger equation (\ref{eq:harmonium})
\begin{equation}
\label{a01}
\left[\op{t}_\ell(r)+v_\mathrm{int}(r;\lambda,\mu)+
\mathcal{R}(r)\right]\varphi_\ell(r;\lambda,\mu)=0,
\end{equation}
where
\begin{equation}
\label{a02}
\op{t}_\ell(r)=-\partial_r^2-\frac{2}{r}\partial_r+
\frac{\ell(\ell+1)}{r^2},
\end{equation}
is the radial part of the kinetic energy operator corresponding to a given
$\ell$,
\begin{equation}
\label{a03}
v_\mathrm{int}(r;\lambda,\mu)=w(r;\mu)+\lambda\bar{w}(r;\mu)
=\frac{(1-\lambda)\,\erf(\mu\,r)+
\lambda}{r},
\end{equation}
is the interaction potential, and
$\mathcal{R}(r)$ is finite at $r=0$. 

After changing variable $r$ to $x=\mu\,r$ and defining
\begin{equation}
\label{a04}
u_\ell(x;\lambda,\mu)=\varphi_\ell(x/\mu;\lambda,\mu)
\end{equation}
eq. (\ref{a01}) becomes
\begin{equation}
\label{a05}
\op{t}_\ell(x)\,u_\ell(x;\lambda,\mu)=\left[\frac{1}{\mu}\,
v_\mathrm{int}(x;\lambda,1)
+\mathcal{O}\left(\mu^{-2}\right)\right]\,u_\ell(x;\lambda,\mu).
\end{equation}

To solve eq. (\ref{a05}) we use the first-order perturbation theory with
perturbation parameter $1/\mu$. We set 
\begin{equation}
\label{a06}
u_\ell(x;\lambda,\mu)=u^{(0)}_\ell(x)+
\frac{1}{\mu}\,u^{(1)}_\ell(x;\lambda)+\mathcal{O}(\mu^{-2}),
\end{equation}
where $u^{(0)}_\ell(x)$ and $u^{(1)}_\ell(x;\lambda)$ are solutions of the 
zeroth and the first order perturbation equations: 
\begin{eqnarray}
\label{a07}
&&\op{t}_\ell(x)\,u^{(0)}_\ell(x)=0,\\
\label{a08} 
&&\op{t}_\ell(x)\,u^{(1)}_\ell(x;\lambda)=F(x;\lambda),
\end{eqnarray}
and
\begin{equation}
\label{a09}
F(x;\lambda)=v_\mathrm{int}(x;\lambda,1)\,u^{(0)}_\ell(x).
\end{equation}
The general solution of the second-order homogeneous differential
equation (\ref{a07}) is a linear combination of its two 
independent solutions: $f_1(x)=x^\ell$, and $f_2(x)=x^{-(l+1)}$: 
\begin{equation}
\label{a10}
u^{(0)}_\ell(x)=\mathcal{A}^{(1)}_\ell\,f_1(x)+
\mathcal{A}_\ell^{(2)}\,f_2(x).
\end{equation}
The inhomogeneous equation (\ref{a08}) is solved in quadratures:
\begin{eqnarray}
\label{a11}
u^{(1)}_\ell(x;\lambda)&=&f_1(x)\,\left[\mathcal{B}^{(1)}_\ell(\lambda)-
\int\,\frac{f_2(x)}{\mathcal{W}(x)}\,F(x;\lambda)\,dx\right]\\
\nonumber
&+&f_2(x)\,\left[\mathcal{B}^{(2)}_\ell(\lambda)+
\int\,\frac{f_1(x)}{\mathcal{W}(x)}\,F(x;\lambda)\,dx\right],
\end{eqnarray}
where
\[
\mathcal{W}(x)=f_1(x)\,\frac{df_2(x)}{dx}-f_2(x)\,\frac{df_1(x)}{dx}=
-\frac{2\ell+1}{x^2}
\]
is the Wronskian of solutions of the homogeneous equation.

According to eqs. (\ref{a03}) and (\ref{a09}),
\begin{equation}
\label{a12}
F(x;\lambda)=(1-\lambda)\,F(x;0)+\lambda\,F(x;1).
\end{equation}
Consequently, from eq. (\ref{a11}) we have
\begin{equation}
\label{a13}
u^{(1)}_\ell(x;\lambda)=(1-\lambda)\,u^{(1)}_\ell(x;0)+
\lambda\,u^{(1)}_\ell(x;1).
\end{equation}
Notice that also
\begin{equation}
\label{a14}
\mathcal{B}^{(i)}_\ell(\lambda)=(1-\lambda)\,\mathcal{B}^{(i)}_\ell(0)+
\lambda\,\mathcal{B}^{(i)}_\ell(1),\;\;\;i=1,2,
\end{equation}
where 
$\mathcal{B}^{(i)}_\ell(0)$ and $\mathcal{B}^{(i)}_\ell(1)$ depend
on neither $\lambda$ nor $\mu$.

The integration constants $\mathcal{A}^{(i)}_\ell$ and
$\mathcal{B}^{(i)}_\ell$, $i=1,2$, are determined from the requirement that
the wave function fulfills the coalescence
conditions\cite{KurNakNak-AQC-16,SavKar-23}.  In particular, neither
$u^{(0)}_\ell(x)$ nor $u^{(1)}_\ell(x;\lambda)$ can be singular at $x=0$. 
The last requirement implies that $\mathcal{A}^{(2)}=0$.

The evaluation of $\lambda=1$ contribution is straightforward.  We have
\begin{equation} 
\label{a15}
u^{(1)}_\ell(x;1)=\frac{1}{2\ell+1}\left[\mathcal{B}^{(1)}_\ell(1)x^\ell+
\frac{\mathcal{B}^{(2)}_\ell(1)}{x^{\ell+1}}\right]+
\frac{\mathcal{A}^{(1)}_\ell\,x^{\ell+1}}{2\ell+2}.  
\end{equation} 
On the other hand, according to eq. (26) of ref.\citenum{SavKar-23}, 
\[
u^{(1)}_\ell(x;1)\,\propto\,\frac{r^{\ell+1}}{2\ell+2} 
\] 
(the Kato's cusp condition\cite{Kat-CPAM-57}).  Therefore, in order to
recover the correct behavior of the Coulomb radial function at small $x$, we
have to set $\mathcal{B}^{(1)}_\ell(1) =\mathcal{B}^{(2)}_\ell(1)=0$. 
Finally, according to eqs.  (\ref{a06}), (\ref{a10}), and (\ref{a15}),
\begin{equation}
\label{a16}
u_\ell(x;1,\mu)=\mathcal{A}^{(1)}_\ell\,x^\ell\,
\left[1+\frac{x}{\mu\,(2\ell+2)}\right]+\mathcal{O}(\mu^{-2}). 
\end{equation}

The second contribution, corresponding to $\lambda=0$, is more difficult to
calculate.  At the limit of large $\mu$, and for sufficiently large $r$,
$\varphi(r;0,\mu)\rightarrow\varphi(r;1,\mu)$~\cite{GorSav-PRA-06} To meet
this condition we have to set $\mathcal{B}^{(1)}_\ell(0)=0$. By fixing
\begin{equation}
\label{a17}
\mathcal{B}^{(2)}_\ell(0)=
-\mathcal{A}^{(1)}_\ell\,\frac{\Gamma(\ell+3/2)}{(2\ell+2)\sqrt{\pi}},
\end{equation}
and using properties of the incomplete gamma function (see, for example,
ref.~\citenum{GraRyz-07}, or
\url{https://functions.wolfram.com/GammaBetaErf/Erf/21/01/02/01/01/01/})
we get an explicit, non-singular, expression  
\begin{equation}
\label{a18}
u^{(1)}_\ell(x;0)=\frac{\mathcal{A}^{(1)}_\ell\,x^\ell}{\sqrt{\pi}}\left[
\frac{\sqrt{\pi}\,x\,\erf(x)}{2\ell+2}+
\frac{\mathrm{e}^{-x^2}}{(2\ell+1)}+
\frac{\Gamma(\ell+3/2)-\Gamma(\ell+3/2,x^2)}{(2\ell+1)(2\ell+2)x^{2\ell+1}}
\right].
\end{equation}
By expanding the right-hand side of eq. (\ref{a18}) to 
a power series of $x$, we get
\begin{equation}
\label{a19}
u^{(1)}_\ell(x;0)=\frac{\mathcal{A}^{(1)}_\ell\,x^\ell}{\sqrt{\pi}}
\sum_{n=0}\,\frac{(-1)^{n+1}\,x^{2n}}{(2n-1)(2n+2\ell+1)\,n!}.
\end{equation}
Alternatively, eq. (\ref{a19}) can be obtained by the expansion of 
$F(x,0)$ and subsequent evaluation of integrals in eq. (\ref{a11}).
Notice, that the integration constants in both approaches are different.
 
For $\ell=0,1,2,\ldots$ eq. (\ref{a18}) can be simplified using the 
recurrence relation given in ref.~\citenum{GraRyz-07}:
\begin{equation}
\Gamma(1/2,x^2)=\sqrt(\pi)\,(1-\erf(x)),\;\;\;\;\;\;
\label{a20}
\Gamma\left(a+1,x^2\right)=a\,\Gamma\left(a,x^2\right)+
x^{2a}\,\mathrm{e}^{-x^2}. 
\end{equation}
\begin{eqnarray}
\label{a21}
u^{(1)}_0(x;0)&=&\frac{\mathcal{A}^{(1)}_\ell}{2}\left[
\left(x+\frac{1}{2\,x}\right)\,\erf(x)+
\frac{\mathrm{e}^{-x^2}}{\sqrt{\pi}}\right],\\
\label{a22}
u^{(1)}_1(x;0)&=&\frac{\mathcal{A}^{(1)}_\ell\,x}{4}\left[
\left(x+\frac{1}{4\,x^3}\right)\,\erf(x)+
\frac{\mathrm{e}^{-x^2}}{\sqrt{\pi}}\left(1-\frac{1}{2\,x^2}
\right)\right],\\
\label{a23}
u^{(1)}_2(x;0)&=&\frac{\mathcal{A}^{(1)}_\ell\,x^2}{6}\left[
\left(x+\frac{3}{8\,x^5}\right)\,\erf(x)+
\frac{\mathrm{e}^{-x^2}}{\sqrt{\pi}}\left(1-\frac{1}{2\,x^2}-
\frac{3}{4\,x^4}\right)\right].
\end{eqnarray}

Combining eqs. (\ref{a06}), (\ref{a10}), (\ref{a13}), (\ref{a15}),
and (\ref{a19}) we get, at the limit of $r\rightarrow{0}$,
\begin{equation}
\label{24}
\varphi_\ell(r;\lambda,\mu)\,\propto\,r^\ell\,\left[1+
\frac{1-\lambda}{\mu\sqrt{\pi}(2\ell+1)}+\frac{\lambda\,r}{2\ell+2}
\right].
\end{equation}

\newpage
\bibliography{biblio}{}
\bibliographystyle{plain}
\end{document}